\documentclass{svproc}

\usepackage{url}

\usepackage{graphicx}
\usepackage{amsmath,amssymb}
\usepackage{booktabs}

\begin{document}
\mainmatter              %

\title{DUNEDepth: Bronchoscopic Depth and Pose Estimation via Hybrid Adaptation of
   a 3D-Vision Foundation Model}
\titlerunning{DUNEDepth: Bronchoscopic Depth and Pose Estimation}
\author{Qi Chang\inst{1} \and Austin Kao\inst{1} \and Ange Lou\inst{2} \and
William E. Higgins\inst{1}$^{\star}$}
\authorrunning{Q. Chang, A. Kao, A. Lou, W. E. Higgins}
\tocauthor{Qi Chang, Austin Kao, Ange Lou, William E. Higgins}
\institute{School of Electrical Engineering and Computer Science,
The Pennsylvania State University, University Park, PA 16802, USA\\
\email{\{qxc62,ack5576,weh2\}@psu.edu}
\and
Vanderbilt University, Nashville, TN 37235, USA\\
\email{lag950516@gmail.com}\\[3pt]
\hspace*{\fill}$^{\star}$\,Corresponding author.\hspace*{\fill}}

\maketitle              %

\begin{abstract}
State-of-the-art bronchoscopy guidance systems and compu\-ter-aided diagnosis methods
have emerged as important tools for early lung cancer detection.  Effective use of
these tools depends on being able to accurately link a patient's 
bronchoscopic airway exam video to their pre-operative 3D chest CT scan.
Unfortunately, the derivation of such links, which include bronchoscope depth and 
camera motion, from the exam video remains an open problem.  Building on recently
proposed 3D-vision foundation models, we propose DUNEDepth, a hybrid architecture
that combines a shared DUNE ViT encoder, a twin MASt3R cross-attention decoder
structure, and a prediction head that includes elements of the MASt3R and Reloc3r
models.  To fine-tune the architecture to our application, we utilize an integrated
hybrid synthetic-supervised/self-supervised training strategy that draws on both
bronchoscopic video and CT-derived virtual bronchoscopic views.  Experiments drawing 
on bronchoscopic video sequences demonstrate that DUNEDepth more accurately derives
depth information and camera motion trajectories than previously proposed
self-supervised methods.

\end{abstract}

\section{Introduction}
\label{sec:intro}

\begin{figure}[!htb]
\centering
\includegraphics[width=0.95\textwidth]{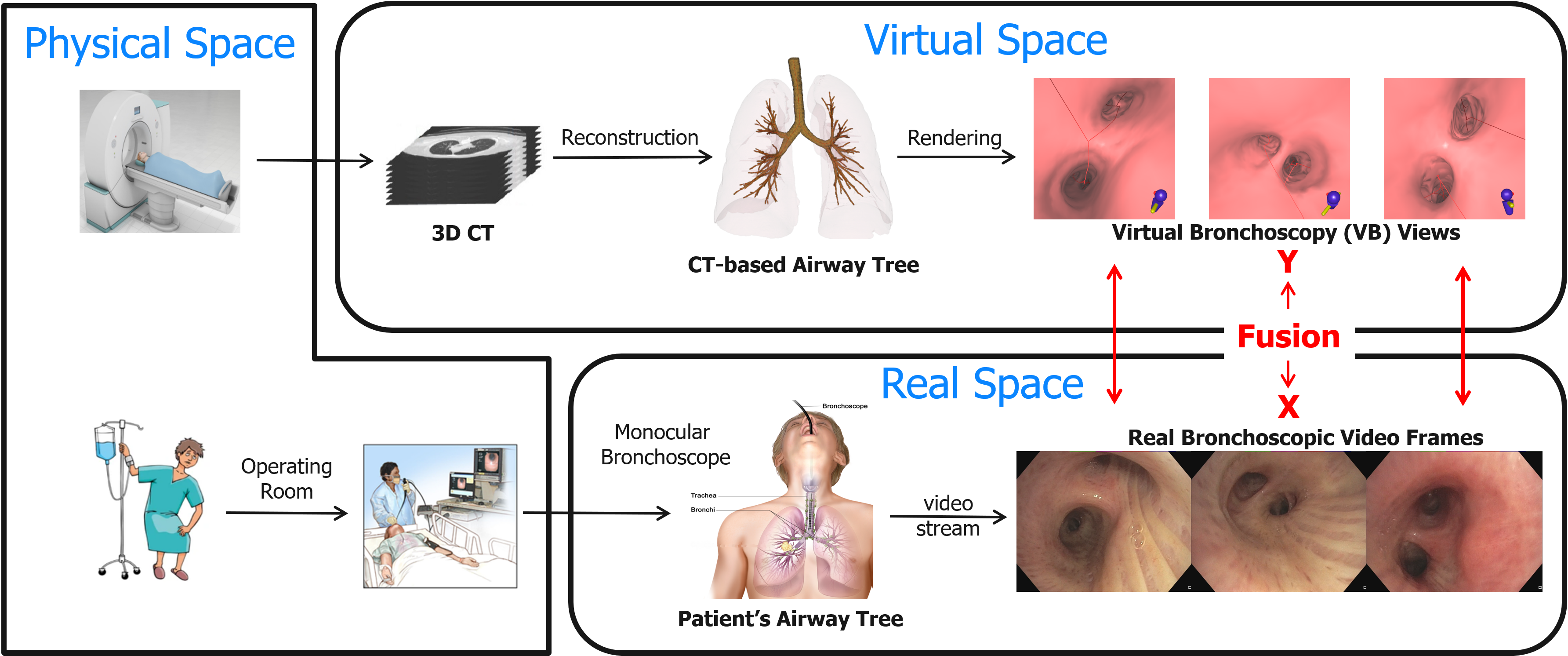}
\caption{Fusion of CT and bronchoscopic video for lung cancer management.}
\label{fig:CTVideo}
\vspace*{-12pt}
\end{figure}

State-of-the-art bronchoscopy guidance systems and computer-aided diagnosis methods 
offer promising tools for early cancer detection in lung cancer patients 
\cite{Gesthalter-ARM2024,Cao-PPIP-Book2025}.  These tools rely on two complementary 
imaging modalities (Figure \ref{fig:CTVideo}): 
1) bronchoscopy, which provides live video imagery for detailed 
visualization of the airway mucosa (walls); and 2) three-dimensional (3D) computed 
tomography (CT) imaging, which gives high-resolution images of the chest and airway 
tree.  By fusing the bronchoscopic video with 3D CT-based airway surface data, 
important diagnostic functions, such as bronchoscope guidance to a suspect peripheral 
nodule, 3D surface reconstruction for mucosal lesion detection, and video generation 
for multimodal lesion analysis can be performed 
\cite{Gesthalter-ARM2024,Cao-PPIP-Book2025,Winiarski-Cancers2025}.

Note, however, that CT-video fusion depends on accurately linking the 
two data sources.  Unfortunately, such direct links do not exist 
because of the known difficulty in acquiring depth and camera ego motion information 
from bronchoscopic video.  In particular, recovering frame-by-frame depth and 
ego motion from monocular bronchoscopic video sequences remains an open problem  
\cite{Shen-RAL2019,Karaoglu-MICCAI2021,Banach-MIA2021}.  Drawing on deep learning 
concepts, the computer vision community has actively addressed this general 
vision problem by devising self-supervised photometric estimators \cite{Godard-ICCV2019,Bian-NeurIPS2019}.  
For example, Monodepth2 introduced the concepts 
of minimum reprojection and automasking, while SC-SfMLearner added scale consistency.  Lastly, 
WS-SfMLearner extended these ideas to surgical videos, which depict open scenes 
exhibiting varied textured surfaces and a wide field of view \cite{Lou-JMI2025}. 

This imaging scenario differs significantly, however, from our more difficult 
situation of imaging the complex airway tree during bronchoscope navigation.  Here, the imaged surfaces 
are confined to an endoluminal tubular structure, with frame-by-frame video variations 
presenting a smooth, slowly varying mucosa, small inter-frame motion, and a photometric 
reprojection signal that carries little information on camera pose 
\cite{Karaoglu-MICCAI2021,Zhang-MICCAI2025}.  Recent bronchoscopy research has 
attempted to address these issues by drawing on a generative adversarial network 
(GAN) to exploit the structural similarity of real video frames and CT-based 
virtual bronchoscopy (simulated video) views during training 
\cite{Shen-RAL2019,Karaoglu-MICCAI2021,Chang-SPIEMI2024a}. None of these works, 
however, supply ego motion supervision.  

On another front, researchers have recently proposed 3D vision foundation 
models for camera pose estimation and relative geometry learning from an image pair. 
These include  DUSt3R, MASt3R, DUNE, and Reloc3r 
\cite{DUSt3R-CVPR2024,MASt3R-ECCV2024,DUNE-CVPR2025,Dong-CVPR2025}.  These works
have motivated several depth and pose estimation approaches for endoscopic video
\cite{Cui-MICCAI2024,Zeinoddin-MICCAI2025,Zhou-MICCAI2025,Shao-CVPR2026}.
No approach, however, considers bronchoscopy or provides full motion 
trajectory information for an endoscopic video stream.

We present a deep learning architecture that enables frame-by-frame depth 
and ego motion prediction, while also providing correctly scaled full-motion trajectory
information for a continuous bronchoscopic video stream.  The architecture, 
referred to as DUNEDepth, adapts the 3D vision foundation model DUNE as an encoder backbone, 
a cross-attention twin MASt3R decoder structure, and a prediction head that provides 
depth, relative camera pose, and trajectory information.  We adapt the architecture 
to the airway tree domain by employing a hybrid CT-video supervision training strategy 
that draws on GAN-based synthetic data and self-supervised inter-frame homography-based 
labels to drive supervised ego motion learning. Section \ref{sec:method} presents our 
architecture, test dataset, and training strategy, Section \ref{sec:exp} presents results 
comparing our approach to existing methods, and Section \ref{sec:conclusion} offers 
concluding remarks.

\vspace*{-6pt}

\section{Methods and Materials}
\label{sec:method}

\subsection{Architecture}
\label{sec:arch}

\begin{figure}[!htb]
\centering
\includegraphics[width=\textwidth]{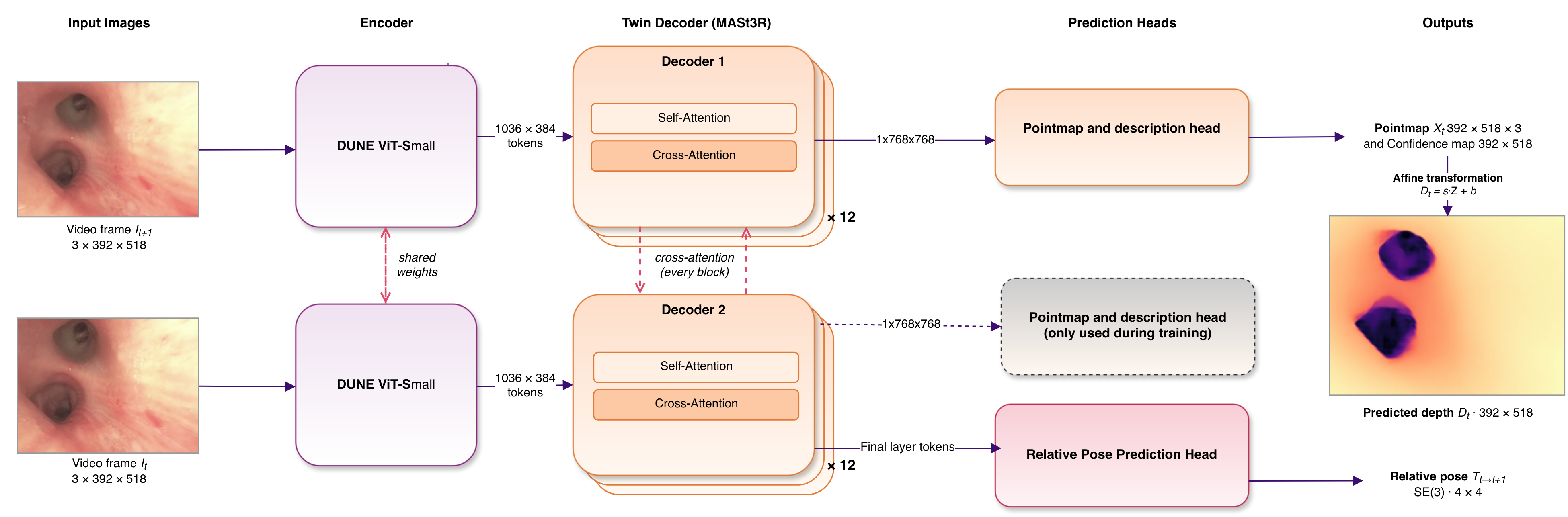}
\caption{DUNEDepth architecture.
}
\label{fig:arch}
\end{figure}

Fig. \ref{fig:arch} depicts the DUNEDepth architecture. It accepts as inputs
a consecutive pair of $392\times518$ video frames $(I_t, I_{t+1})$ that 
have undergone DINOv2 normalization, where the frame pair is 
part of bronchoscopic video sequence 
$\{\ldots, I_{t-1}, I_t, I_{t+1}, I_{t+2}, \ldots\}$.
The inputs pass through a frozen pre-trained weight-shared DUNE ViT-Small 
encoder, which derives DINOv2 semantic and MASt3R geometric features 
\cite{MASt3R-ECCV2024,DUNE-CVPR2025,Oquab-TMLR2024}.
(DUNE = Distilling a UNiversal Encoder; ViT = vision transformer.)
The encoder maps each input frame into a $1036\times384$ token sequence.
Next, a twin MASt3R cross-attention decoder structure processes these token sequences 
over parallel pre-trained MASt3R branches, each consisting of 12 transformer blocks 
(dimension = 768).  Each block in a branch attends to its own sequence using 
self-attention and to the sequence processed by the other branch using cross-attention.
In this way, the two branches exchange geometric information at every level of the
network \cite{MASt3R-ECCV2024}.
Two separate prediction heads next process the decoder output tokens.
First, a pre-trained MASt3R pointmap head uses regression analysis to derive a
camera-space 3D pointmap for each view.  This is used to derive camera-space depth 
$\widehat{D}$ via a learned scalar affine transformation $\widehat{D} = s\cdot Z + b$,
where $Z$ is the unscaled relative depth and parameters $s$ and $b$ are learned scale
and bias values, respectively.
Second, a new Reloc3r-style pose head regresses the SE(3) 4$\times$4 homogeneous 
transformation matrix representing the incremental motion (rotation and 
translation) between frames $I_t$ and $I_{t+1}$ via
Gram-Schmidt orthogonalization \cite{Dong-CVPR2025}.  

Note that the encoder (21.9M parameters) stays frozen to preserve its distilled 
feature knowledge for the weakly textured airways, while the remainder of 
the architecture (329M parameters) is fine-tuned at a low learning rate (Section~\ref{sec:train}).  
Overall, the architecture consists of 350.9M parameters of which 329M are trainable.
Lastly, note that the pose head only reads detached tokens from the last layer of
its source branch, so pose gradients never reach the shared decoder and cannot collapse 
the depth representation during network training.

\vspace*{-6pt}

\subsection{Dataset}
\label{sec:data}

The study uses data from 11 lung cancer patients enrolled under informed 
consent through a research protocol approved by our University’s Institutional Review Board.  
These data include a patient’s 3D chest CT scan and recorded bronchoscopic airway 
exam video stream.  HD-quality Olympus BF-1TH190 bronchoscopes were used for 
the airway exams (30 RGB frames/sec frame rate).  
We created a database consisting of: 1) set $\mathbf{X}$,
consisting of individual video frames derived from the patient airway exams; and 
2) set $\mathbf{Y}$, consisting of virtual bronchoscopy views and corresponding depth
maps from the patient CT scan data.  Figure \ref{fig:CTVideo} depicts $\mathbf{X}$ and $\mathbf{Y}$.
Sets ($\mathbf{X}$, $\mathbf{Y}$) were then 
divided into several other subsets.  First, subset $\mathbf{X_1}$ is comprised 
of 4,209 video frames obtained by sampling every fifth frame of the videos,
while $\mathbf{Y_1}$ consists of 5,283 virtual bronchoscopy views and corresponding 
depth maps.  Note that $\mathbf{X_1}$ and $\mathbf{Y_1}$ do not pair with each other.
We next created paired (registered) video frames with corresponding 
virtual bronchoscopy views and depth maps for 42 different viewpoints. These 
viewpoints were selected to span the 11-patient database and selected to cover 
the various airway bifurcations over both lungs.  32 sets of
these paired images form the validation dataset ($\mathbf{X_2}$, $\mathbf{Y_2}$) 
while the remaining 10 make up a test dataset ($\mathbf{X_3}$, $\mathbf{Y_3}$).  Lastly, 
a subset $\mathbf{X_4}$ contains 130 short contiguous video sequences totaling 9,514 frames 
(a small portion of these are also in set $\mathbf{X_1}$).  For these video
frames, we precomputed frame-to-frame homography rotation pseudo-labels, using a
multi-frame registration method and a RANSAC-based homography estimation pipeline as 
discussed in \cite{Chang-SPIEMI2024a,Byrnes-IEEETBME2019}. The pseudo-labels later help 
supervise ego motion training of DUNEDepth's pose head.

In addition, we created a set of synthetic data to be used during DUNEDepth training.
This was done using subsets $\mathbf{X_1}$ and $\mathbf{Y_1}$ to train a 
CycleGAN-inspired $\mathbf{Y} \rightarrow \mathbf{X}$ domain translator, as
discussed by Chang \cite{Chang-SPIEMI2024a}.  In particular, 
samples from $\mathbf{X_1}$ and $\mathbf{Y_1}$ were used to train a generator and discriminator 
network, which employs the cycle consistency loss of Zhu \textit{et al.} 
\cite{Zhu-ICCV2017}. The generator network tries to transform the input samples from 
$\mathbf{Y}$ into images in $\mathbf{X}$, while the discriminator network evaluates 
the synthetic and real samples to determine which samples are truly from $\mathbf{X}$. 
After completion of the adversarial training, the generator network is used as a 
domain translator to create synthetic video frames $\widehat{\mathbf{X}}$ in the 
real-image domain based on the virtual bronchoscopy views in $\mathbf{Y}$. We
then use these data as pseudo-ground truth during training.

\subsection{Training}
\label{sec:train}

For each step during training, the DUNEDepth architecture selects one of two training strategies: synthetic-supervised training or self-supervised training.
During synthetic-supervised training, we minimized the scale-invariant loss 
$\mathcal{L}_{\rm si}$ 
between predicted depth $\widehat{D}$ and CT-derived depth $D$ \cite{Eigen-NeurIPS2014}:
\begin{equation}
  \mathcal{L}_{\rm si} = \sqrt{\,\textstyle\frac{1}{N}\sum_{i} d_i^{2} - \frac{1}{2}\bigl(\textstyle\frac{1}{N}\sum_{i} d_i\bigr)^{2}\,},\qquad
  d_i = \log D_i - \log \widehat{D}_i ,
\end{equation}
During self-supervised training, we used the loss $\,\mathcal{L}_{\rm photo} + \,\mathcal{L}_{\rm rot}$,
which combines a minimum-reprojection 
photometric term $\mathcal{L}_{\rm photo}$ proposed by Godard with the geodesic SO(3) rotation distance $\mathcal{L}_{\rm rot}$ \cite{Godard-ICCV2019}:

\begin{align}
  \mathcal{L}_{\rm photo}  &= \Bigl\langle \mu \cdot \min_{t' \in \{t-1, t+1\}} \mathrm{\phi}(I_t, \hat{I}_{t'\to t}) \Bigr\rangle + \lambda_{s}\mathcal{L}_{\rm smooth}, \label{eq:Lphoto} \\  \nonumber
            & \mbox{    where  } \mathrm{\phi}(I_a, I_b)= 0.85\cdot\tfrac{1-\mathrm{SSIM}(I_a, I_b)}{2} + 0.15\cdot\bigl|I_a - I_b\bigr|, \\
  \mathcal{L}_{\rm rot}    &= \cos^{-1}\!\left(\mathrm{clip}\!\left(\tfrac{\mathrm{tr}({\bf R}^{\!\top}\!{\bf R})-1}{2},\,-1\!+\!\epsilon,\,1\!-\!\epsilon\right)\right). \label{eq:Lrot}
\end{align}
In \eqref{eq:Lphoto}, $\langle\cdot\rangle$ is the mean over all pixels, $\mathrm{\phi}$ is the 
minimum reprojection error,
$\mathcal{L}_{\rm smooth}$ is the edge-aware smoothness of the mean-normalized depth
$\hat{D}/\langle\hat{D}\rangle$, SSIM is the structural similarity index,
$\lambda_{s}\!=\!10^{-3}$, and mask variable $\mu = 1$.  Regarding \eqref{eq:Lrot},
${\bf R}$ corresponds to the rotation pseudo-labels, tr($\cdot$) = matrix trace, $\epsilon\!=\!10^{-6}$, 
and clip($\cdot$) is the clipping function.
We applied final weights of 1.0 and 0.1 to the synthetic-supervised training loss and the self-supervised training loss, respectively, so that our model prioritizes learning absolute depth prediction.
We sampled between the two training strategies roughly 50/50, and optimized the model with AdamW (weight decay $= 0.01$).
The decoder, the pointmap heads, and the description heads were trained with a learning rate of $10^{-5}$, while the learning rate for the pose prediction head and the affine transform scalars $s$ and $b$ was $10^{-4}$.
Training ran for 10 epochs using batch size = 1.

\section{Results}
\label{sec:exp}

\subsection{Experimental Protocol}
\label{sec:val}

We evaluated depth and trajectory outputs separately. For depth, we compared 
DUNEDepth, Monodepth2, SC-SfMLearner, Depth-Anything-V2, and VGGT 
\cite{Godard-ICCV2019,Bian-NeurIPS2019,Yang-NeurIPS2024,Wang-CVPR2025}.
Monodepth2 and SC-SfMLearner were trained on our data, while Depth-Anything-V2 
and VGGT were used in zero-shot mode with pre-trained weights.
We evaluated depth accuracy using 32-point validation set $(X_2, Y_2)$ and
10-point test set $(X_3, Y_3)$.
We computed L1, L2, and a SSIM-based dissimilarity measure, 
where a lower score implies better performance.
Also, since monocular predicted depth uses an unknown global scale, 
depth predictions must be rescaled before comparison to the ground 
truth.  Hence, we report errors two ways: median-scaled, where a predicted 
depth map is multiplied by the ratio of the median ground-truth depth to the median 
predicted depth \cite{Godard-ICCV2019,Bian-NeurIPS2019}; and non-mapped, where
we scored the raw prediction directly, measuring absolute depth accuracy.
Note that zero-shot models only predict relative depth, so 
absolute depth accuracy cannot be measured.

Next, we used 4 patient video sequences, along with ground-truth CT-video registrations, 
to test trajectory tracking accuracy.  Three sequences were not involved in training, while
the half of the frames in the fourth sequence also appear in the unsupervised training pool.
Before proceeding, let
${\bf P}_i, {\bf Q}_i \in SE(3)$, $i\!=\!0,\dots,N\!-\!1$, denote predicted and reference absolute
poses having camera centers $\mathbf{p}_i, \mathbf{q}_i$. Also, let
$\hat{\mathbf{p}}_i = s\mathbf{R}\mathbf{p}_i + \mathbf{t}$ be the predicted centers
after performing a Sim(3) Umeyama fit given by
$(s,\mathbf{R},\mathbf{t}) = \arg\min_{s,\mathbf{R},\mathbf{t}} \sum_i \lVert
s\mathbf{R}\mathbf{p}_i + \mathbf{t} - \mathbf{q}_i\rVert^2$ \cite{Umeyama-IEEETPAMI1991}. Finally, we have the relative poses
$\Delta {\bf P}_i = {\bf P}_i^{-1}{\bf P}_{i+1}$ and $\Delta {\bf Q}_i = {\bf Q}_i^{-1}{\bf Q}_{i+1}$,
both of which have rotational ($\Delta {\bf R}_i^{\bf P}, \Delta {\bf R}_i^{\bf Q}$) and translational 
($\Delta\mathbf{t}_i^{P}, \Delta\mathbf{t}_i^{Q}$) components.   Given these definitions,
we define the following evaluation metrics: 
absolute trajectory error $\tau_{\rm A}$,  end-point error $\tau_{\rm E}$, and, 
following the RGB-D SLAM benchmark \cite{Sturm-IROS2012},
translation-direction agreement $\rho_{\rm trans}$ and
rotation error $\rho_{\rm rot}$:  %
\begin{align}
   \tau_{\rm A} &= \sqrt{\tfrac{1}{N}\textstyle\sum_{i} \lVert\hat{\mathbf{p}}_i - \mathbf{q}_i\rVert^2}, \qquad
   \tau_{\rm E} = \lVert\hat{\mathbf{p}}_{N-1} - \mathbf{q}_{N-1}\rVert, \\
  \mathrm{\rho_{\rm trans}} &= \tfrac{1}{N-1}\textstyle\sum_{i} \frac{\Delta\mathbf{t}_i^{P}\!\cdot\Delta\mathbf{t}_i^{Q}}{\lVert\Delta\mathbf{t}_i^{P}\rVert\,\lVert\Delta\mathbf{t}_i^{Q}\rVert}, \ 
  \mathrm{\rho_{\rm rot}} = \tfrac{180}{\pi(N-1)}\textstyle\sum_{i} \theta\bigl(\Delta {\bf R}_i^{\bf P}, \Delta {\bf R}_i^{\bf Q}\bigr),
\end{align}
where $\theta({\bf A},{\bf B}) = \cos^{-1}\!\left(\mathrm{clip}\!\left(\tfrac{\mathrm{tr}({\bf A}^{\!\top}{\bf B})-1}{2},\,-1,\,1\right)\right)$.
We compared DUNEDepth's trajectory results against the SC-SfMLearner and zero-shot VGGT models.
To do so, we aggregated each method's source-to-target SE(3) predictions into an absolute trajectory, after which we perform a Sim(3) Umeyama fit to the reference trajectory, which is obtained via CT-video registration.

\begin{table}[!htb]
\vspace*{-12pt}
\centering
\small
\setlength{\tabcolsep}{6pt}
\caption{Depth accuracy results on the 42-point test set  ($({\bf X_2},{\bf Y_2})$, $({\bf X_3},{\bf Y_3})$). 
}
\label{tab:depth}
\begin{tabular}{lccccc}
\toprule
 & \multicolumn{3}{c}{Median-scaled $\downarrow$} & \multicolumn{2}{c}{Non-mapped $\downarrow$} \\
\cmidrule(lr){2-4}\cmidrule(lr){5-6}
Method & L1 & L2 & SSIM & L1 & L2 \\
\midrule
DA-V2$^\dagger$   & 0.090 & 0.023 & 0.269 & --     & --     \\
VGGT-1B$^\dagger$ & 0.088 & 0.020 & 0.254 & --     & --     \\
Monodepth2        & 0.122 & 0.057 & 0.369 & 0.169 & 0.078 \\
SC-SfMLearner     & {\bf \underline{0.067}} & {\bf \underline{0.013}} & {\bf \underline{0.230}} & 0.196 & 0.080 \\
DUNEDepth         & 0.074 & 0.013 & 0.283 & {\bf \underline{0.130}} & {\bf \underline{0.034}} \\
\bottomrule
\end{tabular}
\vspace*{-18pt}
\end{table}

\subsection{Experimental Results}
\label{sec:results}

\begin{figure}[!htb]
\centering
\includegraphics[width=0.95\textwidth]{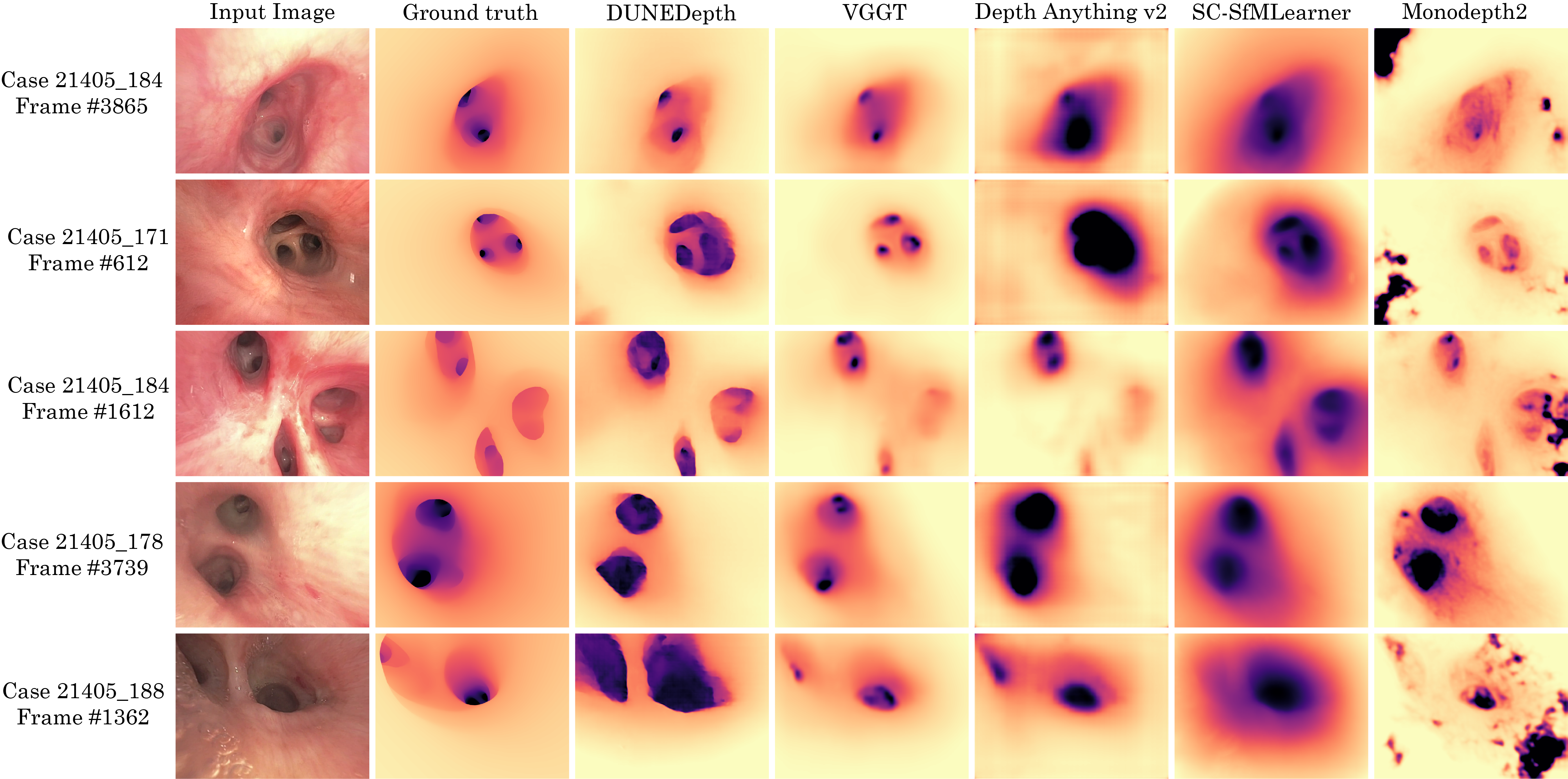}
\vspace*{-6pt}
\caption{Depth-map comparison for three video frames from $({\bf X_3},{\bf Y_3})$ (median-scaled 
relative depth; magma colormap; darker values are deeper).}
\label{fig:depthcmp}
\end{figure}

\begin{figure}[!htb]
\vspace*{-12pt}
    \begin{center}
        \begin{tabular}{cc}
	\includegraphics[width=0.44\textwidth]{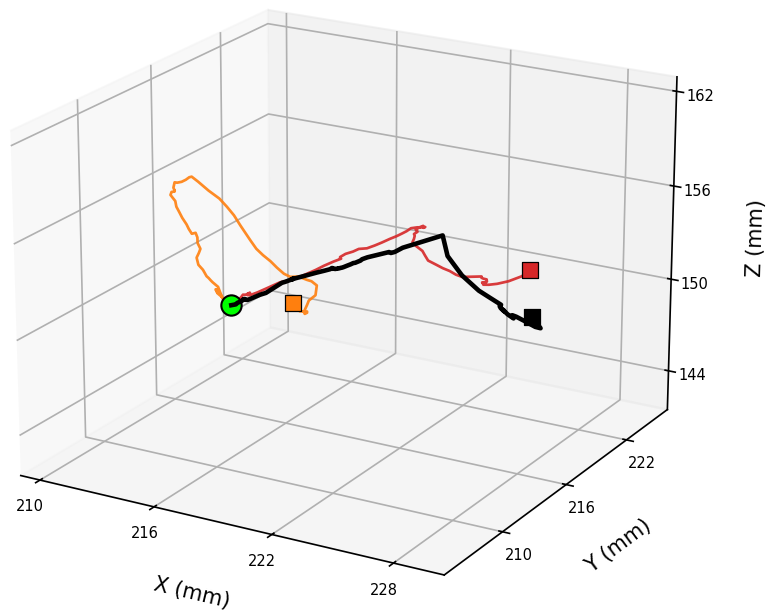} &
	\includegraphics[width=0.44\textwidth]{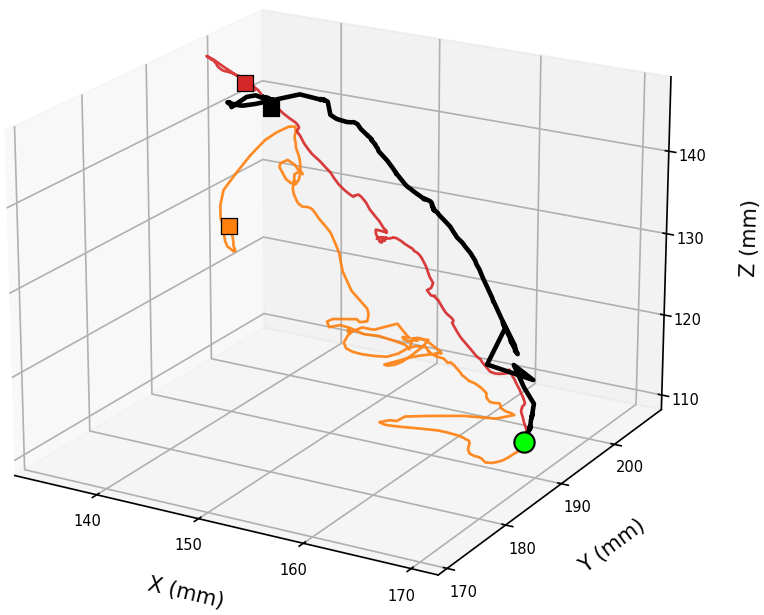}\\
	(a) Sequence B & (b) Sequence C
        \end{tabular}
	\end{center}
\vspace*{-12pt}
    \caption{Trajectory tracking results for two video sequences.
    Black = ground truth trajectory, RED = DUNEDepth, Orange = SC-SfMLearner.
    The green dot is the starting point, while a colored square corresponds to an end point.
    }
    \label{fig:trajcmp}
    \vspace*{-6pt}
\end{figure}

Tables~\ref{tab:depth}-\ref{tab:trajAblation} present numerical results, 
while Figures~\ref{fig:depthcmp} and~\ref{fig:trajcmp} compare individual methods qualitatively.
For all tables, ``$\downarrow$"/``$\uparrow$" means lower/higher values are better,
{\bf \underline{Bold}} indicates the best result for a metric, and
$\dagger$ denotes a zero-shot model.  Lastly, DA-V2 = Depth-Anything-V2.

Per Table~\ref{tab:depth}, DUNEDepth achieves the best absolute depth (non-mapped) 
results.  SC-SfMLearner attains slightly lower median-scaled error, but performs worst 
for absolute depth.  Table~\ref{tab:traj} shows that DUNEDepth  consistently
outperforms or nearly equals all models for the trajectory error $\tau_{\rm A}$ and 
translation-direction agreement $\rho_{\rm trans}$ metrics.  Also, DUNEDepth and 
SC-SfMLearner exhibit small rotation errors $\rho_{\rm rot} \approx 1^o$ 
(not shown).   Depth-map plots of Figure~\ref{fig:depthcmp} corroborate these
findings.  DUNEDepth most closely matches the ground truth, while
SC-SfMLearner gives slightly inferior, less-detailed images.  
The other models give 
manifestly inferior results, with zero-shot DA-V2 and VGGT producing excessively smooth 
depth maps that blur the airway lumen. 
Next, Figure \ref{fig:trajcmp} clearly shows 
DUNEDepth's superior tracking ability over a video sequence, with SC-SfMLearner depicting
significantly more erratic results; VGGT gives essentially unusable tracking results
(not shown in Figure \ref{fig:trajcmp}).
DUNEDepth performed depth and trajectory inference in 0.25 sec and 0.17 sec,
respectively, per input pair on an inexpensive Nvidia RTX 3090 GPU, implying a rate
of 4 outputs/sec (assuming parallel computation).
Overall, DUNEDepth appears to provide the most robust and consistent video tracking.

\begin{table}[!htb]
\vspace*{-8pt}
\centering
\small
\setlength{\tabcolsep}{4pt}
\caption{Trajectory tracking accuracy for four video sequences. Sequences A, B, C, and
D consist of 72, 71, 240, and 331 contiguous frames, respectively. (Note that
CT-video registration introduces $<3$ mm uncertainty in ground-truth poses 
 \cite{Byrnes-IEEETBME2019}.)}
\label{tab:traj}
\begin{tabular}{clccc}
\toprule
Sequence Name & Method & $\tau_{\rm A}$ (mm)$\downarrow$ & $\tau_{\rm E}$ (mm)$\downarrow$ & $\rho_{\rm trans}$ $\uparrow$ \\
\midrule
&SC-SfMLearner     & 4.14 & 5.34 & 0.46 \\
A & VGGT-1B$^\dagger$ & 5.21 & 8.22 & 0.25 \\
 &DUNEDepth         & {\bf \underline{2.22}} & {\bf \underline{2.43}} & {\bf \underline{0.76}} \\
\midrule
&SC-SfMLearner     & 5.61 & {\bf \underline{6.04}} & 0.75 \\
B & VGGT-1B$^\dagger$ & 12.49 & 13.21 & 0.26 \\
  & DUNEDepth         & {\bf \underline{5.22}} & 6.92 & {\bf \underline{0.78}} \\
\midrule
&SC-SfMLearner     & 8.61 & 4.85 &  0.46 \\
C  & VGGT-1B$^\dagger$ & 12.57 & 12.76 & 0.03 \\
   & DUNEDepth         & {\bf \underline{3.84}} & {\bf \underline{3.59}} & {\bf \underline{0.52}} \\
\midrule
&SC-SfMLearner     & 8.13 & {\bf \underline{6.25}} & {\bf \underline{0.22}} \\
D  &VGGT-1B$^\dagger$ & 12.49 & 24.03 & 0.09 \\
   & DUNEDepth         & {\bf \underline{4.32}} & 7.71 & 0.21 \\
\bottomrule
\end{tabular}
\vspace*{-16pt}
\end{table}

\begin{table}[!htb]
\centering
\small
\setlength{\tabcolsep}{6pt}
\caption{Ablation study of DUNEDepth's trajectory tracking accuracy.  Results are based on video sequence D.}
\label{tab:trajAblation}
\begin{tabular}{lcc}
\toprule
Method & $\tau_{\rm A}$ (mm) $\downarrow$ & $\tau_{\rm E}$ (mm) $\downarrow$  \\
\midrule
DUNEDepth         & {\bf \underline{4.32}} & {\bf \underline{7.71}}\\
w/o synthetic, depth images & 5.42 & 8.49 \\
w/o rotation pseudo-labels      & 5.47 & 11.80 \\
\bottomrule
\end{tabular}
\end{table}

Finally, Table~\ref{tab:trajAblation} presents results of an ablation study for DUNEDepth, where
we test the effect of removing different sources of supervision from our training strategy.
We test model performance when synthetic and depth pairs are removed from the training process and when the rotation pseudo-labels are removed from the training process. 
The results show that omitting either element degrades DUNEDepth's trajectory tracking performance.

\vspace*{-4pt}

\section{Conclusion}
\label{sec:conclusion}

To the best of our knowledge, DUNEDepth is the first architecture that estimates
frame-by-frame depth and properly scaled motion trajectory information over
a video sequence from monocular bronchoscopic video.  The architecture builds on 
the lineage of several 3D-vision foundation models that estimate 
camera pose and relative geometry from a pair of video frames---namely, DUNE, 
MASt3R, and Reloc3r \cite{MASt3R-ECCV2024,DUNE-CVPR2025,Dong-CVPR2025}.  We also
proposed a hybrid supervision strategy in a single training loop that combines
synthetic CT-based depth supervision, self-supervised photometric reprojection,
and homography-based motion pseudo-labels to facilitate robust domain learning
of camera depth and ego motion.  In addition, during training, the gradients of 
DUNEDepth's Reloc3r-style pose head remain detached from the shared decoder 
structure, thereby preventing pose supervision from collapsing depth information, 
as can happen with other depth-motion estimation architectures.  Furthermore,
our model draws on a relatively small model (350.9M parameters) that easily runs on 
a single Nvidia RTX 3090 GPU, while surpassing the performance of the 1B-parameter
zero-shot VGGT model \cite{Wang-CVPR2025}.  Results with bronchoscopy videos
derived from lung cancer patients clearly show the accuracy and robustness
of DUNEDepth for depth estimation and motion trajectory calculation.  Lastly,
given that the gastrointestinal tract is also a hollow, tubular organ like
the airway tree, we believe DUNEDepth, coupled with its training strategy, would
readily apply to colonoscopy.  

We note several limitations of DUNEDepth.  First, it could be susceptible to 
under-estimation of residual rotation, as it assumes bronchoscope navigation  
follows the standard video frame rate.  Also, bronchoscopic
video frequently depicts video frames degraded by mucus, bubbles, specular
reflections, bleeding, and blurring arising from rapid motion; such artifacts
could degrade performance.  While our test data  
showed evidence of such degradations at times, we did not explicitly 
evaluate method performance under these conditions.  Lastly, greater robustness
could be gained by training the model on a larger dataset.
Future work will address these issues and consider
application to colonoscopy.  Finally, the method can be applied to improving
existing guided bronchoscopy systems and airway mucosal cancer detection methods.

\vspace*{-6pt}

\bibliographystyle{splncs03_unsrt}
\bibliography{bibtex/MiCAD-WEH2026,bibtex/Chang-June-2026,bibtex/higgins-Aug2024,bibtex/mipl-Aug2024}

\begin{thebibliography}{10}
\providecommand{\url}[1]{\texttt{#1}}
\providecommand{\urlprefix}{URL }

\bibitem{Gesthalter-ARM2024}
Gesthalter, Y.B., Channick, C.L.: Interventional pulmonology: extending the
  breadth of thoracic care. Annual Review Medicine  75(1),  263--276 (2024)

\bibitem{Cao-PPIP-Book2025}
Cao, Z., Ye, B., Zhou, Y., Liu, B.: Artificial intelligence in respiratory
  endoscopy. In: Yarmus, L., et~al. (eds.) Principles and Practice of
  Interventional Pulmonology, pp. 1--16. Springer (2025)

\bibitem{Winiarski-Cancers2025}
Winiarski, S., Radziszewski, M., Korczy{\'n}ski, P., et~al.: Integrating
  artificial intelligence in bronchoscopy and endobronchial ultrasound ({EBUS})
  for lung cancer diagnosis and staging: A comprehensive review. Cancers
  17(17),  2835 (2025)

\bibitem{Shen-RAL2019}
Shen, M., Gu, Y., Liu, N., Yang, G.Z.: Context-aware depth and pose estimation
  for bronchoscopic navigation. IEEE Robot. Autom. Lett.  4(2),  732--739 (Jan
  2019)

\bibitem{Karaoglu-MICCAI2021}
Karaoglu, M.A., Brasch, N., Ladikos, A., et~al.: Adversarial domain feature
  adaptation for bronchoscopic depth estimation. In: Int. Conf. Med. Image
  Comput. Comput. Assist. Interv. pp. 300--310 (Sept 2021)

\bibitem{Banach-MIA2021}
Banach, A., et~al.: Visually navigated bronchoscopy using three
  cycle-consistent generative adversarial network for depth estimation. Med.
  Image Anal.  73(1),  102164--1 --- 102164--37 (2021)

\bibitem{Godard-ICCV2019}
Godard, C., Brostow, G.J., et~al.: Digging into self-supervised monocular depth
  estimation. In: IEEE/CVF Int. Conf. Comput. Vis. pp. 3827--3837 (Oct 2019)

\bibitem{Bian-NeurIPS2019}
Bian, J.W., et~al.: Unsupervised scale-consistent depth and ego-motion learning
  from monocular video. In: Adv. Neural Inf. Process. Syst. vol.~32 (Dec 2019)

\bibitem{Lou-JMI2025}
Lou, A., Noble, J.: {WS}-{S}f{ML}earner: self-supervised monocular depth and
  ego-motion estimation on surgical videos with unknown camera parameters. J.
  Med. Imaging  12(2),  025003 (Mar 2025)

\bibitem{Zhang-MICCAI2025}
Zhang, F.X., Mackute, E., Kasaei, M., Dhaliwal, K., Thomson, R., Khadem, M.:
  {BREA-Depth}: Bronchoscopy realistic airway-geometric depth estimation. In:
  Int. Conf. Med. Image Comput. Comput. Assist. Interv. pp. 75--85 (Sept 2025)

\bibitem{Chang-SPIEMI2024a}
Chang, Q., Higgins, W.E.: Self-supervised monocular depth and ego-motion
  estimation for {CT}-bronchoscopy fusion. In: Proc. SPIE Medical Imaging. vol.
  12928, pp. 47--58. SPIE (2024)

\bibitem{DUSt3R-CVPR2024}
Wang, S., Leroy, V., et~al.: {DUS}t3{R}: Geometric {3D} vision made easy. In:
  IEEE/CVF Conf. Comput. Vis. Patt. Recog. pp. 20697--20709 (June 2024)

\bibitem{MASt3R-ECCV2024}
Leroy, V., Cabon, Y., et~al.: Grounding image matching in 3{D} with {MAS}t3{R}.
  In: Proc. Euro. Conf. Comput. Vis. pp. 71--91. Springer-Verlag (Sept 2024)

\bibitem{DUNE-CVPR2025}
Sariyildiz, M.B., Weinzaepfel, P., Kalantidis, Y., et~al.: {DUNE}: Distilling a
  universal encoder from heterogeneous {2D} and {3D} teachers. In: Proc. IEEE
  Conf. Comput. Vis. Patt. Recog. pp. 30084--30094 (June 2025)

\bibitem{Dong-CVPR2025}
Dong, S., Wang, S., Liu, S., et~al.: {R}eloc3r: Large-scale training of
  relative camera pose regression for generalizable, fast, and accurate visual
  localization. In: Proc. IEEE Conf. Comput. Vis. Patt. Recog. pp. 16739--16752
  (June 2025)

\bibitem{Cui-MICCAI2024}
Cui, B., Islam, M., Ren, H., et~al.: {E}ndo{DAC}: Efficient adapting foundation
  model for self-supervised depth estimation from any endoscopic camera. In:
  Int. Conf. Med. Image Comput. Comput. Assist. Interv. pp. 208--218 (Oct 2024)

\bibitem{Zeinoddin-MICCAI2025}
Sheikh~Zeinoddin, M., et~al.: {E}ndo-{FAS}t3{R}: Endoscopic foundation model
  adaptation for structure from motion. In: Int. Conf. Med. Image Comput.
  Comput. Assist. Interv. pp. 117--126 (Sept 2025)

\bibitem{Zhou-MICCAI2025}
Zhou, Z., Yang, C., Yang, P., et~al.: {E}ndo{DAV}: Depth any video in endoscopy
  with spatiotemporal accuracy. In: Int. Conf. Med. Image Comput. Comput.
  Assist. Interv. pp. 192--201 (Sept 2025)

\bibitem{Shao-CVPR2026}
Shao, S., Zhu, K., Min, Z., et~al.: Depth any endoscopy: Towards
  self-supervised generalizable depth estimation in monocular endoscopy. In:
  Proc. IEEE Conf. Comput. Vis. Patt. Recog. pp. 34126--34137 (June 2026)

\bibitem{Oquab-TMLR2024}
Oquab, M., Darcet, T., Moutakanni, T., et~al.: {DINO}v2: Learning robust visual
  features without supervision. Trans. Mach. Learn. Res.  (2024)

\bibitem{Byrnes-IEEETBME2019}
Byrnes, P.D., Higgins, W.E.: Efficient bronchoscopic video summarization. IEEE
  Trans. Biomed. Eng.  66(3),  848--863 (Mar 2019)

\bibitem{Zhu-ICCV2017}
Zhu, J.Y., Park, T., Isola, P., Efros, A.A.: Unpaired image-to-image
  translation using cycle-consistent adversarial networks. In: Proc. IEEE Int.
  Conf. Compu. Vision. pp. 2223--2232 (2017)

\bibitem{Eigen-NeurIPS2014}
Eigen, D., Puhrsch, C., Fergus, R.: Depth map prediction from a single image
  using a multi-scale deep network. In: Adv. Neural Inf. Process. Syst.
  vol.~27, p. 2366–2374. MIT Press, Cambridge, MA, USA (Dec 2014)

\bibitem{Yang-NeurIPS2024}
Yang, L., Kang, B., Huang, Z., et~al.: Depth {A}nything {V}2. In: Adv. Neural
  Inf. Process. Syst. vol.~37, pp. 21875--21911. Curran Associates, Inc. (Dec
  2024)

\bibitem{Wang-CVPR2025}
Wang, J., Kang, B., et~al.: {VGGT}: Visual geometry grounded transformer. In:
  Proc. IEEE Conf. Comput. Vis. Patt. Recog. pp. 5294--5306 (June 2025)

\bibitem{Umeyama-IEEETPAMI1991}
Umeyama, S.: Least-squares estimation of transformation parameters between two
  point patterns. IEEE Trans. Patt. Anal. Machine Intell.  13(4),  376--380
  (Apr 1991)

\bibitem{Sturm-IROS2012}
Sturm, J., et~al.: A benchmark for the evaluation of {RGB-D} {SLAM} systems.
  In: Proc. IEEE/RSJ Int. Conf. Intell. Robots Syst. pp. 573--580 (Oct 2012)

\end{thebibliography}

\end{document}